\documentclass[10pt,aps,prd,reprint,nofootinbib,floats,floatfix,amsfonts,amssymb,amsmath,preprintnumbers,notitlepage,superscriptaddress]{revtex4-2}

\usepackage[utf8]{inputenc}
\usepackage[margin=1in]{geometry}
\usepackage{graphicx}
\usepackage{xcolor}
\usepackage[colorlinks]{hyperref}
\usepackage[T1]{fontenc}
\usepackage{amsmath, amssymb, amsfonts,amsbsy,amsthm}
\usepackage{academicons,tikz}
\usepackage{subcaption}
\usepackage[normalem]{ulem}

\begin{document}

\title{Prospects for Neutron Star Parameter Estimation using Gravitational Waves from f-modes Associated with Magnetar Flares}

\author{Matthew Ball}
\affiliation{University of Oregon, Eugene, OR 97403, USA}

\author{Raymond Frey}
\affiliation{University of Oregon, Eugene, OR 97403, USA}

\author{Kara Merfeld}
\affiliation{University of Oregon, Eugene, OR 97403, USA}
\affiliation{Texas Tech University, Lubbock, TX 79409, USA}

\date{\today}

\begin{abstract}
Magnetar vibrational modes are theorized to be associated with energetic X-ray flares. Regular searches for gravitational waves from these modes have been performed by Advanced LIGO and Advanced Virgo, with no detections so far. Presently, search results are given in limits on the root-sum-square of the integrated gravitational-wave strain. However, the increased sensitivity of current detectors and the promise of future detectors invite the consideration of more astrophysically motivated methods. We present a framework for augmenting gravitational wave searches to measure or place direct limits on magnetar astrophysical properties in various search scenarios using a set of phenomenological and analytic models.
\end{abstract}

\maketitle

\section{Introduction}
\label{sec:introduction}

Magnetars are a subclass of neutron stars observed to possess magnetic fields with strengths of $10^{14} \: \mathrm{G}$ to $10^{15} \: \mathrm{G}$ (\cite{duncan_thompson_1992}) and may be responsible for the highly energetic emission of X-rays and soft $\gamma$-ray flares. Presently, there are 30 known galactic magnetars\footnote{\url{http://www.physics.mcgill.ca/~pulsar/magnetar/main.html}} found via these (and other) types of emission (\cite{mcgill_magnetar_catalog}). These flares are short duration and believed to be due to interactions between the powerful magnetic field and the exotic matter of the star (\cite{duncan_thompson_1996}).

The detailed internal structure of neutron stars, and magnetars in particular, remains uncertain. Gravitational wave (GW) observations of merging extragalactic binary neutron star systems have already contributed to constraints on the neutron star equation of state (\cite{abbott_gw170817_2017,abbott_gw170817_2018}). GWs emitted from isolated galactic neutron stars would provide a promising new method to explore neutron star astrophysics. The LIGO-Virgo-KAGRA network of GW detectors has carried out a series of searches for GWs associated with magnetar X-ray flares (\cite{04_giant_flare_search,O2magnetar,O3magnetar}). While no detections have resulted, the increasing sensitivity of the detectors begs the question whether such observations are likely in the near future, or with proposed next-generation detectors such as Cosmic Explorer (CE), Einstein Telescope (ET), or the Neutron Star Extreme Matter Observatory (NEMO) (\cite{srivastava_science-driven_2022,Hild_2011,ackley_neutron_2020}).

Unlike binary neutron star mergers, the mechanism(s) for transient GW emission from isolated neutron stars is uncertain. Given the uncertainties, the GW searches to date have made few assumptions regarding specific waveform models. However, a potentially promising scenario associates GW emission in the form of f-modes with the extreme energy release from magnetars observed as giant X-ray flares, with equivalent isotropic energies of $10^{44}$ - $10^{46} \: \mathrm{erg}$ (\cite{mazets_observations_1979,hurley_giant_1999,hurley_exceptionally_2005}). In this paper, we explore the detectability of these f-mode models. As detector sensitivity improves, the probability of detecting a GW signal from a non-compact-binary event increases, and in the event such a signal is measured, complete waveform models are necessary to perform analyses. This is not intended to be the final word on how to analyze these future events, rather a potential metric for gauging astrophysical plausibility of future candidates.

In previous analyses, short-duration, targeted searches use the coherent analysis algorithm X-Pipeline (\cite{sutton_x-pipeline_2010,was_performance_2012,O2magnetar,O3magnetar}). X-Pipeline is a targeted, minimally-modeled, coherent search algorithm. Data from multiple detectors are transformed into time-frequency coherence maps, and the most energetic clusters of pixels which still meet pipeline coherence thresholds after data quality vetoes are recorded. For a nearly monochromatic signal, one might expect a detection to appear as a line of constant-frequency pixels centered on the central frequency of the signal with a narrow bandwidth of only a few Hz. Given the astrophysical uncertainties associated with magnetars and their potential GW emission, we assume here that f-mode emission will be identified by such a minimally-modeled search. However, generic, un-targeted all-sky searches for f-modes associated with pulsar glitches have previously been performed (\cite{lopez_prospects_2022}).

The processes believed to be associated with giant flares in particular are thought to be due to two mechanisms: powerful magneto-hydrodynamic deformations within the interior of star triggered by the cracking of the thin crust of the star and a reconfiguration of the exterior magnetosphere (\cite{levin_excitation_2011}). Magneto-hydrodynamic deformation models predict that a neutron star has fundamental equilibrium states and can suddenly experience a reconfiguration of the internal structure when changing to a different state. This can release an enormous amount of energy potentially emitted through flares or gravitational waves created by the deformation (\cite{ioka2001,corsi_owens_2011}). This deformation can be triggered by a large-scale cracking of the crust (\cite{duncan_thompson_1995,Pacheco1998}), which also serves to provide a channel for the magnetohydrodynamic wave to propagate from inside the star to the exterior magnetosphere (\cite{lyutikov_magnetar_2006,gill_trigger_2010,levin_excitation_2011}).

Theoretical calculations have estimated the total energy released during a giant flare and predict that the GW energy could be high enough to be detectable by near-future detectors by assuming that all available flare energy is released in the GW channel (\cite{corsi_owens_2011}). Other, more detailed calculations give GW energies far below what is detectable with even proposed future GW detectors (\cite{levin_excitation_2011,zink_lasky_kokkotas_2012,ciolfi_poloidal-field_2012}).

The vibrational modes excited within the neutron star are believed to be predominately fluid pressure modes. The fundamental `f-mode' is the lowest order ($l=m=2$) pressure mode, allowing for the possible generation of gravitational waves (\cite{mcdermott_nonradial_1988}) expected to occur in the approximate range of 1 to 3 kHz (\cite{duncan_thompson_1995,Pacheco1998}). These excitations are expected to be short-lived. Pressure oscillations within the star are expected to couple with the core and primarily decay through GW emission, causing the oscillations to last for $\mathcal{O}(0.1-10 \: \mathrm{s})$ (\cite{levin_qpos_2006}).

This paper is organized as follows: In Section \ref{sec:fitting_functions}, we discuss the various fitting functions and analytic models that can allow us to create a fuller picture of the gravitational waves from an f-mode. In Section \ref{sec:usage} we will explore how this picture can be used to evaluate the plausibility of f-mode emission for detection candidates identified in an umodeled search. We note that in Section \ref{subsec:full_detection}, we perform a similar Bayesian analysis to that done by \cite{clark_evidence-based_2007}. However, we analyze over a larger frequency/damping time space and focus more on direct parameter estimation and future detector sensitivity. This section is also similar to recent work done by \cite{pradhan_constraining_2023}; however, we focus on EOS-independent relations for f-modes from X-ray flares and consider recovery in the context of an initial, unmodeled search result. Finally, in Section \ref{sec:future_detetors}, we will explore how these measurements stand to be improved with future detectors.

\section{Methods}
\label{sec:fitting_functions}

If we consider a magnetar with a monochromatic oscillatory mode at frequency $\nu_{GW}$ spontaneously excited at time $t=0$ and decaying with a timescale $\tau_{GW}$, then, following the logic of \cite{Echeverria1989} and \cite{finn1992}, we can express the gravitational wave emission of an f-mode as a damped sinusoid of the form in Equation \ref{eq:h(t)}

\begin{equation}
h(t) = 
    \begin{cases}
        0 & \text{for } t < 0\\
        h_0 e^{-t/\tau_{GW}} \sin{\left(2\pi \nu_{GW}t\right)} & \text{for } t \geq 0
    \end{cases}
    \label{eq:h(t)}
\end{equation}

The emission is then characterized by an amplitude $h_0$, a characteristic frequency, and a characteristic damping time. While no model currently exists that properly describes the parameters (mass, radius, magnetic field strength) of this system due to the unknown neutron star equation-of-state (EOS), attempts have been made to form EOS-independent fitting functions to approximate these parameters. These fitting functions utilize numerical simulations to map such physical parameters to measurable quantities.

\subsection{Frequency and Damping Time Fitting Functions}
\label{subsec:freq_damp_fitting_functions}

A variety of attempts have been made to compute relationships between GW observables like frequency and damping time and physical properties like mass and radius. These typically involve performing numerical simulations at a variety of different EOS's and finding a best-fit function that relates these properties (see eg. \cite{andersson_gravitational_1996, kokkotas_inverse_2001, benhar_gravitational_2004, tsui_universality_2005, andersson_gravitational_2011, gaertig_gravitational_2011, doneva_gravitational_2013, kruger_kokkotas-2020}). We choose to consider the fitting functions of \cite{anderson_kokkotas-1998} as most other calculations either arrive at similar results or they expand to include fast rotation -- which we do not consider as most known magnetars are nowhere near this regime and the rotating fits converge to the non-rotating models in the zero spin limit. These fitting functions are shown in Equations \ref{eq:freq} and \ref{eq:damp} where $\overline{M} = M/1.4 M_{\odot}$ and $\overline{R} = R/10 \: \mathrm{km}$.

\begin{equation}
\nu_{GW}(kHz)\approx0.78+1.635\left(\frac{\overline{M}}{\overline{R}^3}\right)^{1/2}
\label{eq:freq}
\end{equation}

\begin{equation}
\frac{1}{\tau(s)}\approx \frac{\overline{M}^3}{\overline{R}^4}\left[22.85-14.65\left(\frac{\overline{M}}{\overline{R}}\right)\right]
\label{eq:damp}
\end{equation}

Assuming a mass range of [$0.8 \; M_{\odot}$, $2.5 \; M_{\odot}$] and a radius range of [$8 \; \mathrm{km}$, $20 \; \mathrm{km}$], this model allows f-modes to exist in the frequency range [1.22 kHz, 3.33 kHz] and damping time range [$0.087 \; \mathrm{s}$, $10 \; \mathrm{s}$]. This includes cuts on the parameter space requiring the damping time and frequency to be positive (as the algebraic fitting functions would technically allow negative values) and for the mass and radius to satisfy Buchdahl's bound (\cite{buchdahl_general_1959}). Since the algebraic fitting functions would technically allow damping times as high as $2500 \; \mathrm{s}$, we place a cut when sampling this parameter space that the damping time must be less than $10 \: \mathrm{s}$ as calculations predict damping times below this limit (\cite{levin_qpos_2006}) and the numerical simulations the fitting functions were drawn from do not give damping times greater than this (\cite{anderson_kokkotas-1998}).

We note here that these relations do not give unique solutions for all frequency/damping times. A small part of parameter space (roughly between $2000-2700\: \mathrm{Hz}$ and damping times below around $0.23\: \mathrm{s}$), the same frequency/damping time can correspond to multiple mass/radius values. Due to the dependence of the amplitude on these parameters, one of these modes can be eliminated to some extent by constraining the magnetic field if that information is known.

\subsection{Theoretically motivated amplitude model}
\label{subsec:amplitude_model}
Another function can describe the amplitude of the oscillation. By modeling the oscillation as a reconfiguration of the internal magnetic field triggered by some impulse, the energy of an f-mode oscillation can be expressed as (\cite{levin_excitation_2011})

\begin{equation}
    E_f = \frac{\epsilon_0^2 c^4}{128\pi^2} \frac{\alpha_n^2 B^4 R^4}{q_M M \nu^2}
    \label{eq:amplitude_alt}
\end{equation}

where $\epsilon_0$ is vacuum permittivity. Here, $\alpha_n$ represents the scale of the magnetic field reconfiguration, with $\alpha_n=1$ corresponding to a complete reconfiguration of the entire field, and the quantity $q_M M$ is the effective mass of the oscillating mode, which is dependent on the specific equation of state, the mass, and the radius. The magnetic field strength $B$ here is a characteristic value of the surface magnetic field of the magnetar\footnote{This may not be the same magnetic field as estimated from spindown measurements which assume a dipole field subject to magnetic braking. The spindown measurements are where field strengths of order $10^{14} \: \mathrm{G}$ come from, but this surface field value may be different.}.

We connect this energy to the peak amplitude in Equation \ref{eq:h(t)} via the gravitational wave luminosity as a function of source distance given in equation 21 of \cite{owen_how_2010} when including a damping term, and integrating from $t=0$ to $t=\infty$ which gives Equation \ref{eq:h0} as in (\cite{ho_gravitational_2020}).

\begin{equation}
    h_0 = \frac{1}{\pi d \nu_{GW}} \left(\frac{5G}{c^3}\frac{E_{GW}}{\tau_{GW}}\right)^{1/2}
    \label{eq:h0}
\end{equation}

If we assume that all of the energy of the f-mode is radiated as gravitational waves (\cite{thorne_nonradial_1969,detweiler_variational_1975}), then we can let $E_f = E_{GW}$ and combine Equations \ref{eq:amplitude_alt} and \ref{eq:h0} to get a full description of the wave amplitude.

Given that $\alpha_n$, $q_M$, and the magnetic field strength $B$ are degenerate with each other, we can improve things by fixing $B$ to the dipole magnetic field strength estimated from measurements of the spindown of the source magnetar (\cite{mcgill_magnetar_catalog}); however, this may not be the best choice since the surface field and the dipole field may be quite different. Nonetheless, if we let these fields be comparable, this constraint would allow us to incorporate information on the source magnetar into analyses and focus on the ratio of the reconfiguration scale $\alpha_n$ to the effective mode mass $q_M$ which is related to the equation of state (\cite{levin_excitation_2011}). In the limit of a complete magnetic field reconfiguration $(\alpha_n \rightarrow 1)$ and a mode mass in the range $0.02<\alpha_{n}<0.07$, the amplitude of an f-mode oscillation computed using the above model agrees with the results of numerical simulations for oscillations due to magnetic re-configurations performed by \cite{lasky2012}.

\section{Applications in GW Scenarios}
\label{sec:usage}

Previous work has predicted that galactic magnetar X-ray flares could excite f-modes with energies detectable by second-generation GW detectors (\cite{corsi_owens_2011}); however, other calculations (\cite{zink_lasky_kokkotas_2012}) suggest that second-current detectors are unlikely to detect an f-mode from a galactic flare. Nevertheless, in the event search pipelines do find an astrophysical candidate, it is desirable to connect the putative candidate with astrophysical source parameters. We can construct three levels of analysis for different degrees of candidate strength. In the event the candidate is of high enough amplitude to perform full parameter estimation, we need to know what level of precision source parameters can be extracted and what thresholds need to be met to perform further analyses. In the event a candidate event is detected, but too weak for full parameter estimation, we need some method of determining source parameters without full parameter estimation. Finally, in the event a search returns no GW candidates, can we place any interesting limits on astrophysical parameters?

\subsection{Usage in a Full Detection}
\label{subsec:full_detection}

A full detection would take the form of a search result with sufficiently high probability of an astrophysical source. In this case, we consider how to extract astrophysical information from the data. In particular, we can make statements on the consistency of the signal with an f-mode with reasonable astrophysical parameters.

If we include an inclination term in our model to differentiate plus and cross polarizations, we can use the fitting functions as full model for full parameter estimation. For example, using the \textsc{Bilby} Markov-Chain Monte Carlo (MCMC) infrastructure (\cite{ashton_bilby_2019}) and the \textsc{LALInference} package (\cite{Veitch_LAL_2015}), we can perform simulated injections to attempt to identify the range of parameter space where a signal injected into noise can be recovered. A range of signals was generated using the model presented above and coherently injected into simulated, Gaussian, detector noise using packages in \textsc{LALInference} using noise sensitivity estimates for the third observing run (\cite{buikema_sensitivity_2020}). Source orientation and sky location were chosen to maximize the signal strength in both of the LIGO detectors at a distance of $10 \: \mathrm{kpc}$, corresponding roughly to the center of the galaxy -- and to SGR 1935+2154, an interesting magnetar since it has emitted fast radio bursts as well as X-ray flares (see \cite{lyman_fast_2022}). We then used \textsc{Bilby} to attempt to recover the parameters of the injected signals.

Injections were performed over a somewhat physically motivated parameter space, with masses ranging from 1.0 to 2.6 $M_{\odot}$, radii ranging from 10 to 14 km, and surface magnetic field strengths ranging from $10^{14}$ to $10^{17} \: \mathrm{G}$ -- which is beyond what is generally considered possible, but may not be that optimistic in light of more recent measurements (see \cite{sobyanin_ultraslow_2023}). All injections were performed assuming a fixed distance of $10 \: \mathrm{kpc}$ -- a distance consistent with the center of the galaxy, an optimal inclination angle, and a fixed sky location, chosen to maximize the antenna response in both LIGO detectors. This represents a best-case detection scenario such that the limiting factors of recovery would be intrinsic source parameters. This is also consistent with previous GW searches (\cite{O2magnetar,O3magnetar}).

For our simulated recoveries, the prior distributions for the core parameters were as follows: mass was uniform from 0.5 to 3.0 $M_{\odot}$, radius was uniform from 8 to 16 km, and the magnetic field was log uniform from $10^{13}$ to $10^{17}  \: \mathrm{G}$. We also assumed a full magnetic field reconfiguration such that $\alpha_n = 1$.

A general quantity used to reflect the amplitude of a signal is $h_{\mathrm{rss}}$ (Equation \ref{eq:hrss}). This quantity can be directly compared to the noise ASD for the detectors.

\begin{equation}
    h_{\mathrm{rss}}^2 = \int_{-\infty}^{\infty}\left|h_+(f)\right|^2+\left|h_{\times}(f)\right|^2 \mathrm{df}
    \label{eq:hrss}
\end{equation}

\begin{figure}
    \centering
    \includegraphics[width=\linewidth]{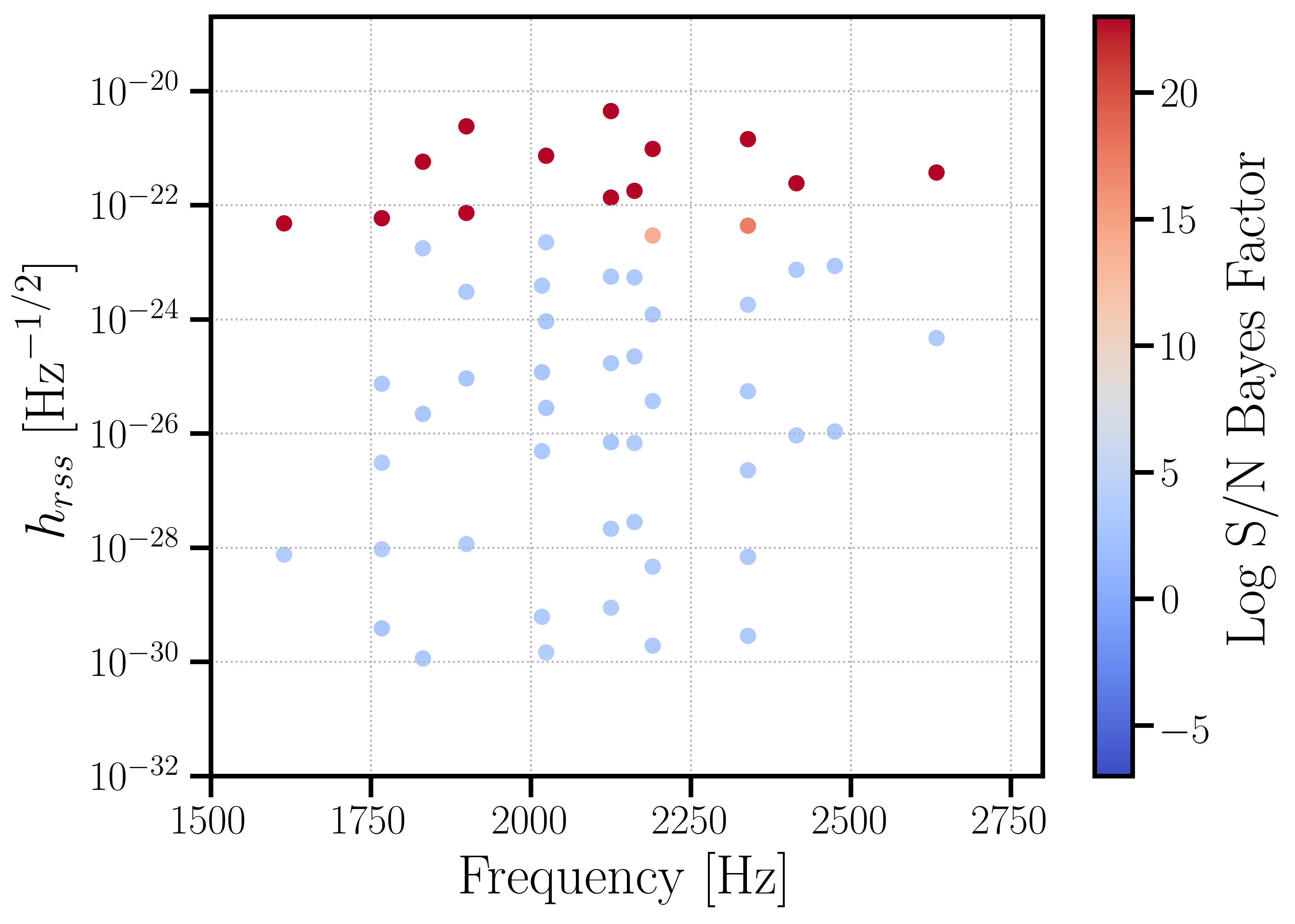}
    \caption{Frequency space representation of injection recoverability for the LIGO H1-L1 network at O3 sensitivity (\protect\cite{buikema_sensitivity_2020}). Each point represents one injection, identified by the frequency and $h_{\mathrm{rss}}$ of the signal and the signal-to-noise log Bayes factor of the recovery. The $h_{\mathrm{rss}}$ threshold for recovery is around $3\times10^{-23} \: \mathrm{Hz}^{-1/2}$ and frequency independent. The color scheme of the log Bayes factor is truncated at 22 to emphasize the recovery point despite many injections recovered with log Bayes factors of $>10^6$}
    \label{fig:freq_vs_hrss}
\end{figure}

\begin{figure}
    \centering
    \includegraphics[width=\linewidth]{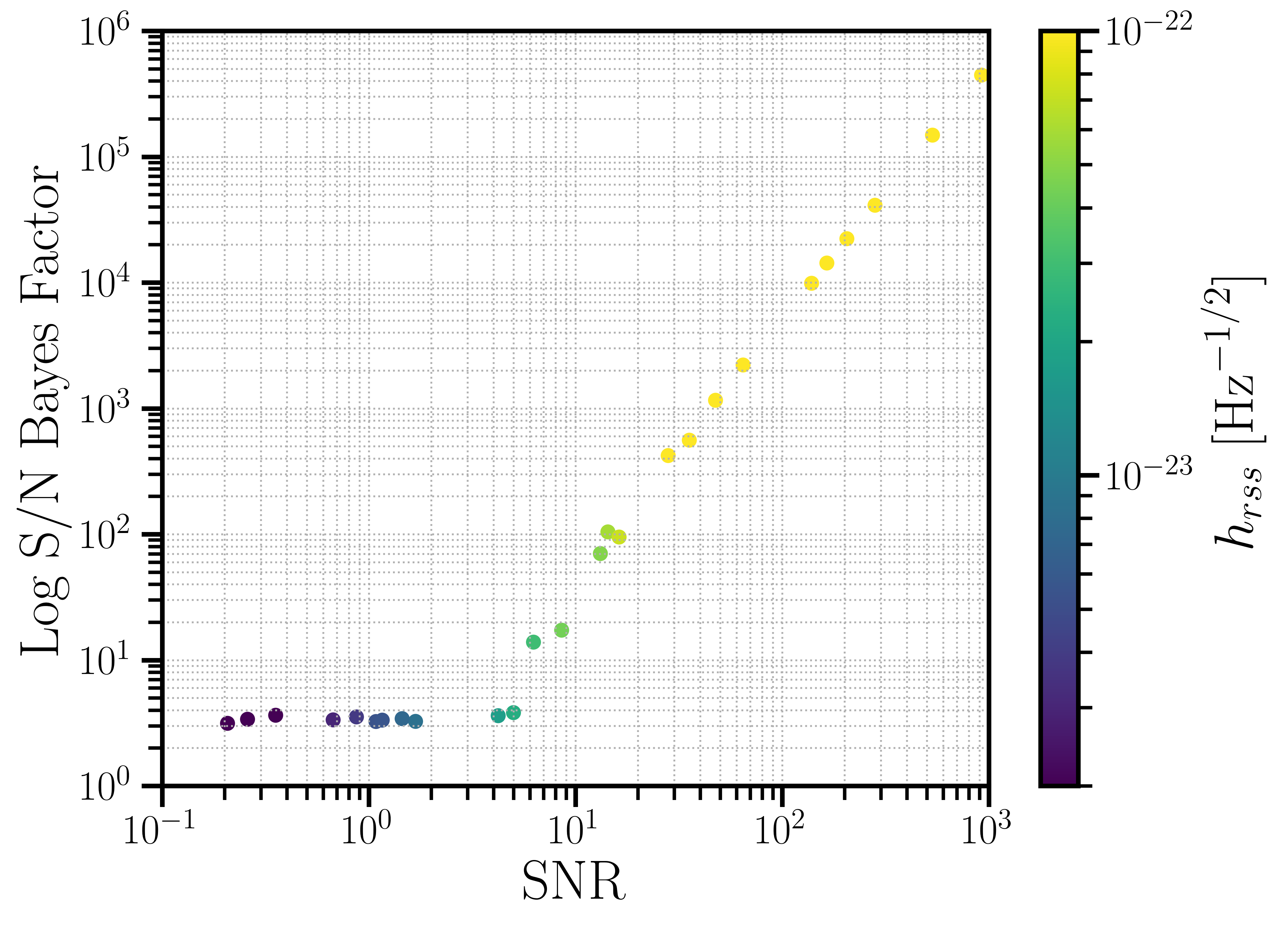}
    \caption{Matched-Filter SNR vs log Bayes factor for a range of signals injected into simulated Gaussian noise for the LIGO H1-L1 network at O3 sensitivity (\protect\cite{buikema_sensitivity_2020}). Injected signals ranged in mass, radius, and magnetic field strength, but given a known distance of 10 kpc. The log Bayes factor sharply increases at an SNR of 8.}
    \label{fig:full_detection}
\end{figure}

Each injection recovery produced a log Bayes factor \footnote{All log Bayes factors reported in this paper are natural log.} for how well the model was preferred over noise. The set of injections, organized by the frequency of the damped sinusoid and the $h_{\mathrm{rss}}$ of the waveform, and colored by this log Bayes factor, are shown in Figure \ref{fig:freq_vs_hrss}. The color scale here was fixed to be centered on a log Bayes factor of 8. This value of 8 is typically chosen as the threshold of `strong evidence' (\cite{jeffreys_theory_1998}). In this regime, injections that were recovered are colored red, and those that were not recovered are blue. This gives an approximate $h_{\mathrm{rss}}$ threshold of $3\times10^{-23}  \: \mathrm{Hz}^{-1/2}$. Casting this into a frequency-independent picture, Figure \ref{fig:full_detection} shows how the signal-to-noise log Bayes factor directly compares to the SNR of the injections and shows the clear recoverability threshold SNR of 8. This is similar to the thresholds estimated by \cite{clark_evidence-based_2007}.

In terms of recovery accuracy, above an SNR of 8, the injected mass and radius were within the $90\%$ confidence limit (CL) but the width of that limit was frequency-dependent due to detector sensitivity. For example, one injection with $(\mathrm{M},\mathrm{R})=(1.0 M_{\odot}, 14 \mathrm{km})$ at a frequency of about $1600 \: \mathrm{Hz}$ with a magnetic field of $30\times10^{15} \mathrm{G}$ had an SNR of 13. This injection was recovered with a $90\%$ confidence interval $(\mathrm{M},\mathrm{R})=( 1.13_{-0.18}^{+0.18}M_{\odot}, 14.9_{-1.0}^{+0.9}\mathrm{km})$. Another injection with $(\mathrm{M},\mathrm{R})=(1.8 M_{\odot}, 14 \mathrm{km})$ at a frequency of about $1900 \: \mathrm{Hz}$ with a magnetic field of $30\times10^{15} \mathrm{G}$ had an SNR of 16. This injection was recovered with $(\mathrm{M},\mathrm{R})=( 1.9_{-0.4}^{+0.5}M_{\odot}, 14.7_{-1.4}^{+1.1}\mathrm{km})$. The difference in SNR here is due to the effect of the mass and the frequency on the amplitude.

Since real detector noise contains non-Gaussian elements, the use of Gaussian noise here means this represents an optimistic scenario, and any attempts to recover a signal from real data will likely see a higher $h_{\mathrm{rss}}$ threshold for recovery. The analysis done here is meant to form a baseline for later examination of predicted sensitivity curves of future detectors (see Section \ref{sec:future_detetors}).

\subsection{Usage in a Marginal Detection}
\label{subsec:marginal_detection}

A marginal detection may take the form of a search result where a pixel cluster is identified by X-Pipeline as a candidate, but the probability of an astrophysical source is near or just below the threshold for proper detection. In this case, full parameter estimation is unlikely and analyses are limited to what could be allowed given the time-frequency window identified by the search. Estimates of the signal morphology can be made with an unmodeled, coherent signal reconstruction algorithm like \textsc{BayesWave} (\cite{cornish_bayeswave_2015}) which attempts to reconstruct a signal coherent between different detectors using Morlet–Gabor wavelets.

\begin{figure}
\centering
\includegraphics[width=0.8\linewidth]{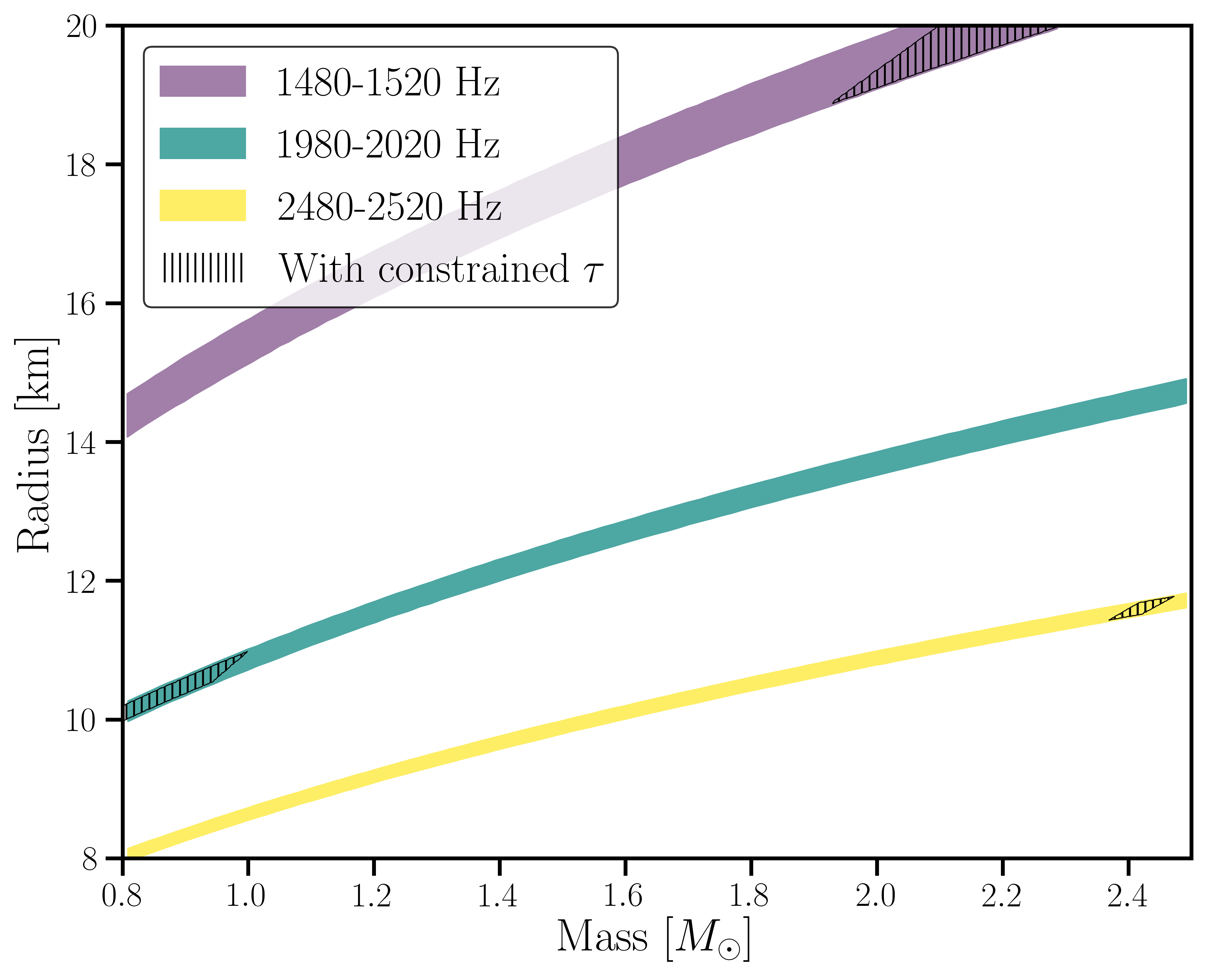}
\caption{Allowed parameter space for a signal given frequency limits from a targeted, unmodeled search. Narrow frequency windows around central frequencies of 1500, 2000, and 2500 Hz show clear mass-radius relations that can be extracted from a potential marginal detection. Marked regions are when we constrain the damping time to $0.3\:\mathrm{s} < \tau <0.4\:\mathrm{s}$.}
\label{fig:marginal_detection}
\end{figure}

Under a hypothetical analysis, we can use Monte Carlo techniques to generate random samples based on physically motivated source distributions and check if these could match the time-frequency window of the marginal candidate. Without amplitude restrictions (as search algorithms rely nearly entirely on coherence statistics), the frequency window of the marginal candidate can be used to place constraints on what astrophysical parameters would allow such a candidate. Figure \ref{fig:marginal_detection} shows what the mass-radius restriction looks like for frequency windows centered on different central frequencies of 1.5, 2.0, and 2.5 kHz. Including damping time constraints could allow analyses to focus in on where in the mass-radius band the source could exist, but determining damping time from the time duration of the trigger is not an obvious task. The triggers show the precise amount of time that the loudest $1\%$ of pixels were clustered together, which is not necessarily the decay time of a damped sinusoid. We can make weaker estimates for an analysis, such as limiting the damping time to be less than twice the reported duration. Additionally, reconstruction algorithms like \textsc{BayesWave} can be used to estimate the damping time from the reconstruction. If there is a candidate duration of several tens of seconds, it is unlikely that it could be adequately explained by this model.

\subsection{Usage in a Non-Detection}
\label{subsec:non_detection}

In the event that there are no confident candidates returned by the search, astrophysical information can still be extracted from the non-detection. Previous searches have placed limits on the $h_{\mathrm{rss}}$ of the signal via injections at different frequencies. In our hypothetical analysis, we can use these limits to constrain the amplitude of a signal and sample our physically motivated parameter space with the only restriction being that the $h_{\mathrm{rss}}$ at some frequency is below the upper limits set by the search and a maximum damping time of $10 \: \mathrm{s}$. Since the source distance is likely known, this primarily can place limits on the three degenerate terms: magnetic field strength $B$, scale of magnetic reconfiguration $\alpha_n$, and oscillation mode mass fraction $q_M$.

\begin{figure}
    \includegraphics[width=0.9\linewidth]{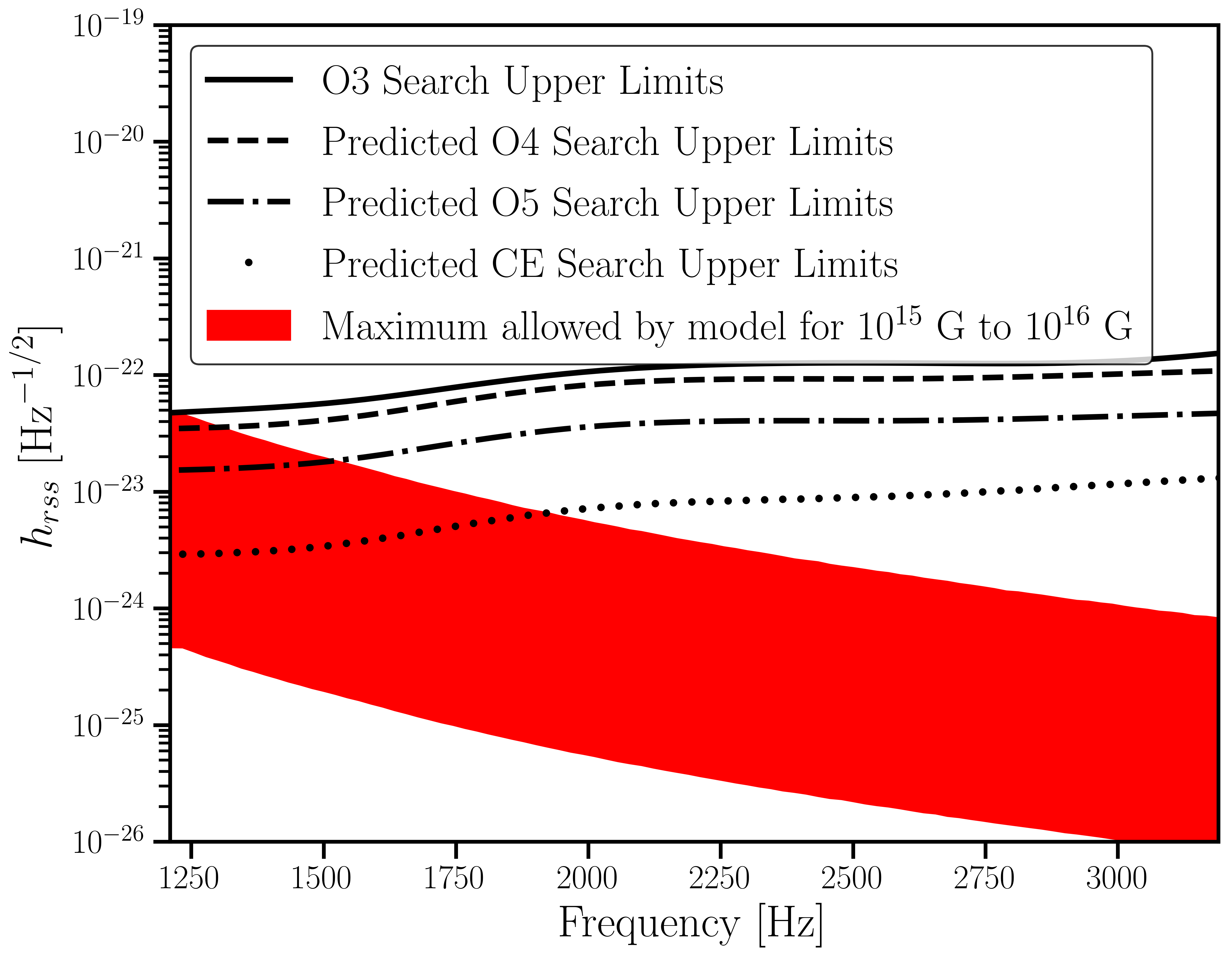}
    \caption{Example results using non-detection $h_{\mathrm{rss}}$ $50 \%$ upper limits from one target of the O3 magnetar search with an estimated source distance of $3.8 \: \mathrm{kpc}$, assuming a magnetic field of between $10^{15}$ to $10^{16} \: \mathrm{G}$ (far greater than the estimated field strength of the original source). The red region shows the maximum allowed $h_{\mathrm{rss}}$ limits when forcing the mode mass fraction to be 0.046 for the SLy equation of state (as calculated in \protect\cite{levin_excitation_2011}) and the magnetic field reconfiguration scale to be 1 for a complete field reconfiguration. Projections of these upper limits onto future detector sensitivities are also shown.}
    \label{fig:non-detection_limits}
\end{figure}

If the dipole magnetic field estimated from spindown measurements is assumed to be equal to the surface magnetic field that modulates the amplitude of the f-mode, some of the degeneracy of the model can be removed. In this case, $\alpha_n$ and $q_M$ may have limits placed. Further constraints can be placed by restricting the mode mass to values calculated for modern equations of state allowing for limits to be placed on the scale of magnetic reconfiguration.

Figure \ref{fig:non-detection_limits} shows an example of a comparison we can make in the event of a non-detection in a hypothetical analysis. Upper limits for the $h_{\mathrm{rss}}$ for a non-detection from a flare potentially associated with 1 RXS J170849 (as done in \cite{O3magnetar}) at a distance of $3.8 \; \mathrm{kpc}$  are compared to what the model would allow for a magnetic field of $10^{15}$ to $10^{16} \; \mathrm{G}$. The maximum allowed $h_{\mathrm{rss}}$ meets the search upper limits at low frequencies in this example, which might suggest that a low-mass, high-radius neutron star (required for a lower frequency) is right at the threshold of detectability. The upper limits are also scaled to the predicted sensitivity of future detectors (\cite{buikema_sensitivity_2020,preO4_noise,Aplus_GWINC,srivastava_science-driven_2022}). LIGO A+ (O5) and 3rd generation detectors like Cosmic Explorer will have the potential to place strong exclusion limits on the allowed parameter space in the event of non-detections. This is a very optimistic scenario, as of the 30 known magnetars, only one has a magnetic field in this range, and of the magnetars with known distances, only about six are comparable to or nearer than this distance (\cite{mcgill_magnetar_catalog}).

\subsection{Something Unexpected}
\label{subsec:unexpected}

In the event a candidate is present that exists outside of the parameter space allowed by f-mode models, a few scenarios can be considered. A first approach for an unexpected candidate would be to attempt to extract a general waveform using an unmodeled reconstruction algorithm like \textsc{BayesWave} (\cite{cornish_bayeswave_2015}). With a candidate reconstruction, the $h_{\mathrm{rss}}$ and the gravitatitonal-wave energy assuming isotropic emission from the source trigger distance can be calculated. From this energy estimate, energetics arguments could be made about whether or not the candidate is astrophysically plausible. \cite{corsi_owens_2011} estimated the limits of energies that an internal magnetic reconfiguration could reach. The energy of an f-mode is limited by the energy released by changes in magnetic deformation, so a magnetic reconfiguration that drives an f-mode cannot expend more energy than available in the reservoir. If the energy of a candidate exceeds what can possibly be stored in the deformation reservoir, it is unlikely to be physical.

If a short-duration, monochromatic candidate is found that occurs outside of the allowed frequency space of this model, it may be possible to make arguments that the candidate was due to a different vibrational mode of the neutron star. While the f-mode models considered here are believed to be most likely for GW emission (\cite{mcdermott_nonradial_1988,detweiler_variational_1975}), frequencies above a few kHz are expected to be where one might find higher order pressure modes (or p-modes) (using the notation of \cite{anderson_kokkotas-1998}). Frequencies below $1 \: \mathrm{kHz}$ are where Alfv\'en modes are predicted that are potentially associated with quasi-periodic oscillations observed in electromagnetic measurements. However, these are complex and have not been fully modeled (\cite{lasky2012}). Crustal and shear modes are also hypothesized to occur in this range (\cite{duncan_global_1998,piro_shear_2005}), as are g-modes with buoyancy as the restoring force (\cite{reisenegger_new_1992}), though these are unlikely to decay via the emission GWs as efficiently as the f-mode (\cite{mcdermott_nonradial_1988}).

Another possibility is a candidate with appropriate frequency and duration but with an amplitude exceeding what the model allows. In the O3 magnetar search, the most significant candidate was a short burst of bright coherent pixels between 1560 and 1608 Hz lasting 63 ms (\cite{O3magnetar}). The frequency and duration fall right in the region expected for an f-mode, but the burst candidate was excluded as an astrophysical candidate for data quality reasons. However, it is instructive to consider what it would imply if the burst candidate had not been a data quality artifact.

While the frequency constraint of 1560 - 1608 Hz fits well within the model discussed here, a damping time of 63 ms is below the minimum damping time allowed by the model. If we consider the case where less than one complete damping time was detected by a search pipeline, then we could have a damping time greater than this value. The source magnetar for the triggering flare was SGR 1935+2154 which has an estimated dipole magnetic field strength of $2.2\times 10^{14} \: \mathrm{G}$ (\cite{israel_discovery_2016}) and an estimated distance of $9.0\pm 2.5  \: \mathrm{kpc}$ (\cite{zhong_distance_2020}). If we sample this physically motivated part of parameter space, we get a maximum $h_{\mathrm{rss}}$ of $\sim 7\times 10^{-27} \: \mathrm{Hz}^{-1/2}$ and a maximum isotropic energy of $1.5\times 10^{31} \: \mathrm{ergs}$ at $9.0 \: \mathrm{kpc}$. This is well below the upper limit $h_{\mathrm{rss}}$ set by the search at $1.3\times10^{-22} \: \mathrm{Hz}^{-1/2}$ at 1600 Hz (\cite{O3magnetar}). If an astrophysically motivated model with the same time-frequency shape as a candidate has an $h_{\mathrm{rss}}$ over 4 orders of magnitude below the candidate, the candidate is either not astrophysical, or not described by the f-mode model. The discrepancy here demonstrates the ability of the model to help vet potential candidates with a more astrophysical motivation.

Another way of thinking about the burst candidate is that to get the astrophysically motivated model to match the amplitude set by the upper limits of the search, the magnetic field strength would need to be increased by about 2 orders of magnitude and the distance reduced to about $4 \: \mathrm{kpc}$. While not astrophysically feasible as an f-mode event for the targeted magnetar (SGR 1935+2154 at $9.0\pm2.5 \: \mathrm{kpc}$), the burst candidate would not be unfeasible within the context of this f-mode model for a magnetar at the distance of 1 RXS J170849 ($3.8 \: \mathrm{kpc}$) but with a magnetic field of $\sim 10^{16} \: \mathrm{G}$, which is at the very upper limit of what magnetars could reach in some estimates (\cite{sobyanin_ultraslow_2023}). Likewise, if it were due to magnetar with $10^{15} \: \mathrm{G}$ at a distance of $10 \: \mathrm{pc}$ -- which is about 100 times closer than the nearest magnetar, it could also be consistent with the model.

\section{Discussion and Conclusions}
\label{sec:future_detetors}

The relationship between SNR and log Bayes factor shown in Section \ref{subsec:full_detection} allows us to estimate sensitivity to potential signals for future detectors. Using predicted noise thresholds (\cite{buikema_sensitivity_2020,Aplus_GWINC,srivastava_science-driven_2022}), we can quantify how future detectors will improve searches for these signals.

\begin{figure}
    \centering

    \includegraphics[width=0.9\linewidth]{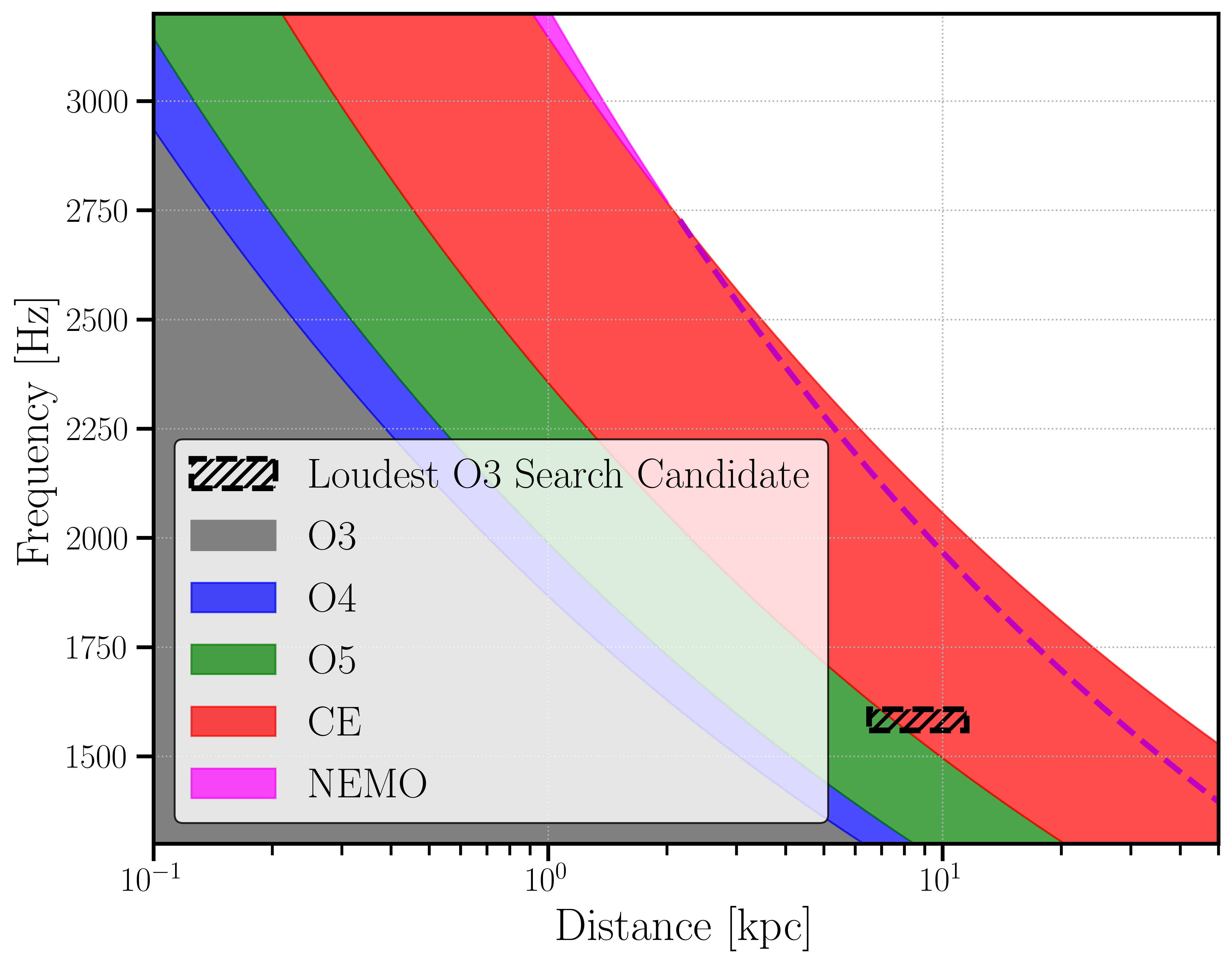}
    \label{fig:future_sensitivities_3sigma}
    
    \caption{Regions of the distance-frequency parameter space that future detectors could be sensitive to assuming a complete magnetic field reconfiguration and source magnetic field strength of $10^{16} \: \mathrm{G}$. The parameter space used in section \ref{sec:usage} was sampled and binned in signal frequency and distance. The fraction of samples in each bin with an expected SNR above a threshold of 3 was calculated, and the region where at least $10\%$ meet this threshold are shaded. The strong frequency dependence is primarily due to the frequency dependence of the sensitivity curves of the detectors. Potential detections can be compared to such a calculation to vet the plausibility of a real signal. The loudest candidate from the O3 search is marked as a dotted box and falls squarely within the Cosmic Explorer and NEMO detection thresholds. This suggests that for such a candidate to have been from a magnetar f-mode with realistic source parameters, the detector would have needed to be at third generation detector sensitivity. Due to the considerable similarity in this frequency regime between Cosmic Explorer and the Einstein Telescope, only the Cosmic Explorer estimate is shown to represent these third generation detectors (\protect\cite{srivastava_science-driven_2022, Hild_2011}). It is also worth noting that NEMO has very similar sensitivity to these signals as the third generation detectors in this frequency space (\protect\cite{ackley_neutron_2020}).}
    \label{fig:future_sensitivities}
\end{figure}

Figure \ref{fig:full_detection} shows that at a matched-filter SNR of about 8, the log Bayes factor spikes to nearly 10 for injection recovery. We can compute what signal parameters would give a particular SNR using future detector noise curve estimates to investigate the prospects for future observing runs. Figure \ref{fig:future_sensitivities} shows the frequency-distance parameter space for a giant flare-induced f-mode with a fixed surface magnetic field strength of $10^{16} \: \mathrm{G}$ and magnetic reconfiguration scale of $1$, allowing the mass, radius, and mode mass to vary, where at least 10$\%$ of samples in a distance-frequency bin have an SNR greater than 3. This $3 \: \sigma$ threshold was chosen to correspond to the p-value threshold used for potentially interesting X-pipeline clusters to follow-up on, which is about 1\% (\cite{sutton_x-pipeline_2010}). A giant flare from a high-magnetic field magnetar is similar to what was observed from SGR 1806-20 -- though it has a slightly weaker field strength (\cite{hurley_exceptionally_2005}). The strong frequency dependence is primarily due to the frequency dependence of the sensitivity curves of the detectors. This parameter space represents what searches might expect to see in the event of a possible candidate: a known source distance, and a frequency space for the candidate. By comparing a potential candidate with the potential detection ranges for different detectors, we can make astrophysical arguments about why a signal may or may not be an f-mode. The loudest candidate from the O3 search (\cite{O3magnetar}) described in Section \ref{subsec:unexpected} is included for comparison.

Here, we have demonstrated applications for a more complete description of the gravitational wave emission from neutron star f-modes associated with magnetar X-ray flares. We have shown that in simulated LIGO noise, full MCMC reconstruction may be possible above a matched-filter SNR of 8 which occurs at an $h_{\mathrm{rss}}$ of around $3\times 10^{-23} \: \mathrm{Hz}^{-1/2}$. We have shown the limits that could be placed on a mass-radius relation in the event of a marginal detection with just frequency information. We have demonstrated how non-detections may be able to constrain the astrophysical parameter space as sensitivity improves. We have also explored how this model can be used to vet potential candidates using astrophysically motivated predictions. Lastly, we have shown how the available search space will improve with the increased sensitivity of future detectors.

\section*{Acknowledgements}

This material is based upon work supported by NSF’s LIGO Laboratory which is a major facility fully funded by the National Science Foundation. The authors acknowledge access to computational resources provided by the LIGO Laboratory supported by National Science Foundation Grants PHY-0757058 and PHY-0823459.
The authors would like to thank Paul Lasky for useful comments and discussions. 
This paper has been given LIGO DCC number P2300341.
MB and RF are supported by UO NSF grant PHY-1912604. During this project, KM was supported by UO NSF grant PHY-1912604.

\section*{Data Availability}

The data underlying this article are available publicly in the Gravitational Wave Open Science Center (https://gwosc.org/). The derived data used in figures will be shared on reasonable request to the corresponding author.

\bibliographystyle{mnras}
\bibliography{references}

\label{lastpage}
\end{document}